\documentclass[%
reprint,
aps,
pra,
amsmath,
amssymb,
superscriptaddress,
]{revtex4-1}

\usepackage[utf8]{inputenc}
\usepackage[T1]{fontenc}

\usepackage{natbib}
\usepackage{amsmath}
\usepackage{graphicx}
\usepackage{xcolor}
\usepackage{xspace}
\usepackage{hyperref}

\graphicspath{figures/}

\newcommand{\dD}{\delta_D}
\newcommand{\dLS}{\delta_\mathrm{LS}}
\newcommand{\tramsey}{T_\mathrm{R}}
\newcommand{\mdD}{\left\langle\dD\right\rangle}

\begin{document}

\title{Velocity-dependent phase shift in a light-pulse atom interferometer}

\author{Léo Morel}
\email{leo.morel@lkb.upmc.fr}
\affiliation{Laboratoire Kastler Brossel, Sorbonne Université, CNRS, ENS-PSL
University, Collège de France, 4 place Jussieu, 75005 Paris}
\author{Zhibin Yao}
\affiliation{Laboratoire Kastler Brossel, Sorbonne Université, CNRS, ENS-PSL
University, Collège de France, 4 place Jussieu, 75005 Paris}
\author{Pierre Cladé}
\affiliation{Laboratoire Kastler Brossel, Sorbonne Université, CNRS, ENS-PSL
University, Collège de France, 4 place Jussieu, 75005 Paris}
\author{Saïda Guellati-Khelifa}
\affiliation{Laboratoire Kastler Brossel, Sorbonne Université, CNRS, ENS-PSL
University, Collège de France, 4 place Jussieu, 75005 Paris}
\affiliation{Conservatoire National des Arts est Métiers, 292 rue Saint
Martin, 75003 Paris, France}

\begin{abstract}
    Atom interferometry relies on the separation and recombination of atom wavepackets. When the two paths overlap perfectly at the end of the
interferometer, the phase is insensitive to the atomic velocity 
distribution.
Here, we show that, when the separation and recombination is performed using a Raman transition there is a displacement of the atomic wavepacket due 
to a phase shift during light pulses.
Because of the variation of the laser intensity seen by the atoms, there 
is an imperfect cancellation of these displacements. The observation of 
a velocity-dependent phase shift on the interferometer is the signature 
of this effect, which has been modeled. Thanks to the signature we have 
identified, we are able to compensate for this effect by applying laser 
power ramps during the interferometer to mitigate intensity variations.
\end{abstract}

\maketitle

\section{Introdution}

Light-pulses used to perform two-photon Raman transitions or Bragg diffraction enable to coherently separate and recombine atomic wave functions \cite{kasevich_atomic_1991, giltner_atom_1995}. 
Based on such pulses, atom interferometers have been used in various applications such as the measurement of the Newtonian gravitational constant\cite{fixler_atom_2007, rosi2014}, of the fine structure constant \cite{bouchendira_new_2011, parker_measurement_2018}, or have served to test the equivalence principle \cite{dimopoulos2007, schlippert2014, barrett_correlative_2015, rosi2017}. Achieving the accuracy required for these applications relies on careful designs. Most notably,
the time sequence of the light pulses that operates the atom interferometer is always arranged in a symmetric way so that the two interfering atomic wavepackets overlap at the end. In this case, the initial velocity of the atoms does not affect the phase of the atom interferometer. This condition is similar to that of an optical interferometer working in white light, when the delay between the two arms is identical.

Usually, the computation of the phase includes only the phases imprinted by the lasers on the atoms that perform the transition \cite{storey_feynman_1994}. Nevertheless, atoms remaining in the initial state undergo an additional phase shift that results in a displacement of the central position of the atomic wavepacket. The magnitude of this displacement depends on the laser intensity perceived by the atoms. If the atoms perceive a constant laser intensity over the interferometric sequence, this displacement is the same for both arms and thus the interferometer remains closed. 

However, in experiments, the laser beams that drive the optical transitions are spatially
finite with an intensity profile that is usually 
Gaussian. Moreover, the atoms that are interrogated are not at zero temperature 
and possess a transverse velocity (orthogonal to the beams propagation direction). Hence, 
the intensity that an atom perceives along the interferometric sequence varies: both the transverse velocity of the atoms and the intensity profile of the Raman laser beams restore sensitivity to the longitudinal velocity distribution. 
The sensitivity of atom interferometers is usually enhanced by increasing their duration. However, in such a configuration the expansion of
the atomic cloud is larger, therefore the bias associated with this effect compromises the gain in sensitivity.
We evaluate the corresponding phase shift to a few tens of mrad, which does not affect the contrast of the interferometer.

The effect we report here has already been studied in~\cite{gillot_limits_2016}. Using the formalism of the sensitivity function, the authors 
investigated how this phase shift sets a limit to the symmetry of the $\pi/2-\pi-\pi/2$ atom gravimeter. Notably, their work focused on the impact of this parasitic phase shift on the accuracy and long-term stability of the gravity measurement.

In this paper, we present a more general approach.  We have theoretically modeled this phase shift for a Raman transition, which enables to evaluate its contribution in other interferometer geometries. We then study experimentally the behavior of this phase shift as a function of the atomic mean velocity and propose a method to mitigate its contribution to the total phase shift.
We used the experimental setup designed to the
measurement of $h/m$ \cite{bouchendira_new_2011} where the atom interferometer is realized using Raman transitions and the $^{87}$Rb atom cloud is produced in an optical molasses with a 
$1/e$ velocity distribution width of 2 cm/s ($T\approx4\ \mu$K).

Because of light shifts during Raman transitions, the variation of intensity during the interferometer also induces a related phase shift, which is coupled to the wavepacket displacements that we study here. In this paper, we compute a general formula that takes into account both effects. Furthermore, the method of compensation that we present also limits the impact of the light shifts.

The paper is organized as follows: we start by presenting a theoretical model to evaluate this displacement of the atomic wavepacket and to provide a general formula of the phase shift. Then, we present our experiment and the protocol we set up to observe this phenomenon. Finally, we show that we mitigate the
related systematic bias by compensating for the variation of the laser power perceived by the atoms during the sequence of the atom interferometer.

\section{Calculation of the wave-packet displacement}

\begin{figure}
    \centering
    \includegraphics{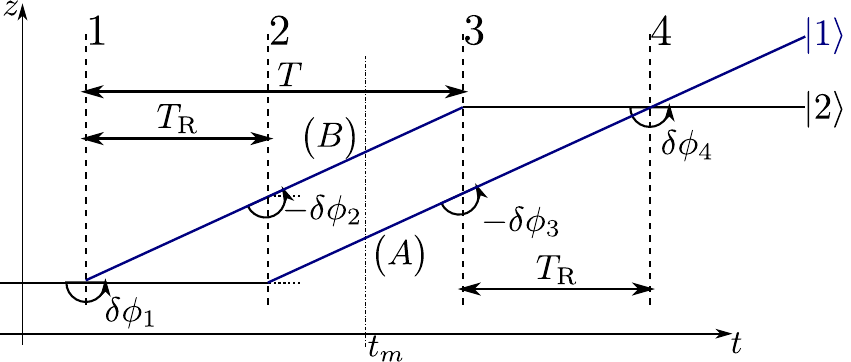}
    \caption{Schematic of the Ramsey-Bordé interferometer implemented in our
    experiment. At the beginning of the interferometer the atoms are in
    $\left|2\right\rangle$. After the first Ramsey sequence of duration
    $\tramsey$, the atoms that remain in this state (dashed lines) are
    removed by a resonant blow-away pulse. After a duration $T$, we apply a
    second Ramsey sequence of same duration to close the interferometer.}
    \label{fig:ramsey_borde_dphi}
\end{figure}

In our experiment, we use a so-called Ramsey-Bordé
interferometer \cite{borde_atomic_1989} which consists of four $\pi/2$ Raman pulses \cite{moler_theoretical_1992} driven by counterpropagating lasers and arranged in two
identical Ramsey sequences of duration $\tramsey$ separated by a duration $T$ (see figure (\ref{fig:ramsey_borde_dphi})).

Two-photon Raman transitions couple the two hyperfine levels of the fundamental state. This coupling depends on the effective  Rabi frequency $\Omega$, the light pulse duration $\tau$ and the detuning of the transition that is the sum of two terms: the Doppler shift
$\dD=\vec{k}_R\cdot\vec{v}$, where $\vec{k}_R$ is the effective wavevector of the lasers that drive the Raman transition and $\vec{v}$ the velocity of the atom, and the
differential light shift $\dLS$ induced by the Raman coupling. 

The Hamiltonian of the system $H=H_0+V_c$ includes the free evolution Hamiltonian $H_0$ and coupling $V_c$, which write
\begin{align}
    H_0 &= \frac{\hbar\dD}{2}\sigma_z 
    = \hbar\begin{pmatrix} \dD/2 & 0\\0 & -\dD/2 \end{pmatrix}
    \label{eq:ham_free}\\
    V_c &= \frac{\hbar\Omega}{2}\sigma_x + \frac{\hbar\dLS}{2}\sigma_z
    = \hbar\begin{pmatrix} \dLS/2 & \Omega/2\\\Omega/2 & -\dLS/2 \end{pmatrix}\label{eq:ham_coupl}
\end{align}
We first assume for simplicity that $\dLS=0$. The case $\dLS\neq0$ will be investigated later. 

In the following, we treat Raman transitions  using the evolution operator. From equations 
(\ref{eq:ham_free}) and (\ref{eq:ham_coupl}), we obtain the 
evolution operators associated to $H_0$ and $H$:
\begin{align}
    U_0(\tau) &= \begin{pmatrix} \exp\left(-i\dD\tau/2\right) & 0\\
        0 & \exp\left(i\dD\tau/2\right) \end{pmatrix}\\
    U(\tau) &= \cos\left(\Omega_e\tau/2\right)1
    - i\sin\left(\Omega_e\tau/2\right)\frac{\left(\Omega\sigma_x
    + \dD\sigma_z\right)}{\Omega_e}\text{,}
    \label{eq:full_evol_no_ls}
\end{align}where $\Omega_e^2=\Omega^2+\dD^2$ is the generalized Rabi
frequency. With the evolution operator
written in equation (\ref{eq:full_evol_no_ls}) we observe that the phases of
the off-diagonal elements are independent of the velocity, which is not the case for
the diagonal elements.

In order to understand how this dependency in $\dD$ translates into a
position shift, we follow the treatment of \cite{jannin_phase_2015}. We write
the two states $\left|1\right\rangle$ and $\left|2\right\rangle$ and assume
that the initial state is a one-dimensional (along the Raman wavevector
direction) Gaussian wavepacket in state $\left|2\right\rangle$ that writes in
momentum space $\widetilde{\Psi}(p, 0)$. For simplicity, here we assume that
the initial momentum distribution is centered at zero. At a time $\tau$ after the beginning of the Raman transition, the state writes:
\begin{align}
    \widetilde{\Psi}(p, \tau) = \left(\cos\left(\frac{\Omega_e \tau}{2}\right) +
    i\frac{\dD}{\Omega_e}\sin\left(\frac{\Omega_e \tau}{2}\right)\right)\widetilde{\Psi}(p,0)\text{.}
    \label{eq:psi_tau}
\end{align} The mean position of the wavepacket $\left\langle x(\tau)\right\rangle$ is given by
\begin{align}
    \left\langle x(\tau)\right\rangle = \left. -\hbar\frac{\partial\phi(\tau)}{\partial p}\right|_{p\rightarrow0}
\end{align} where $\phi(\tau)$ is the
phase of the wavepacket at time $\tau$. Using equation (\ref{eq:psi_tau}), it writes:
\begin{align}
    \left\langle x(\tau)\right\rangle = -v_R\frac{\tan\left(\frac{\Omega\tau
    }{2}\right)}{\Omega}\text{,}
    \label{eq:displacement_raman}
\end{align}where $v_R=\hbar k_R/m$ is the atomic recoil associated with a
Raman transition.

For a single pulse , we observe that this displacement depends on the
Rabi frequency of the transition. During an interferometric sequence, the cloud expands in
beams with a Gaussian intensity profile and the average Rabi frequency decreases. As a consequence, the displacement does not 
compensate along the interferometer, which is not closed. Thus, this
effect induces a sensitivity to the initial velocity of the atoms.

Using equation
(\ref{eq:displacement_raman}), we compute the order of magnitude of the spread between the two wavepackets at the output of a Ramsey-Bordé interferometer that we implement in our experiment 
(figure (\ref{fig:ramsey_borde_dphi})). For this computation, we assume a linear decrease of the Rabi 
frequency with fractional rate $\beta$:
\begin{align}
    \Omega(t)=\Omega(1-\beta(t-t_m))\mathrm{,}
    \label{eq:linear_rabi_var}
\end{align} 
where $t_m$ is the middle point of the interferometer (see figure 
(\ref{fig:ramsey_borde_dphi})). We obtain:
\begin{align}
    \Delta\!\left\langle x\right\rangle = (\pi-2)\frac{v_RT}{\Omega}\beta
    \text{.}
    \label{eq:displacement_rb}
\end{align} The spread between the wavepackets increases with
the duration $T$ between the Ramsey sequences, and is reduced with the
duration of the $\pi/2$ pulses. For typical parameters of our experiment 
$\Omega=2\pi\times5\ $kHz, $T=30\ $ms and $\beta=2\times10^{-3}\ $ms$^{-1}$
(decrease of $\Omega$ by 10\% over 50 ms), we find a displacement of 13 nm. This computation based on formula (\ref{eq:displacement_raman}) allows us to compute the displacement for a narrow velocity
distribution centered on the Raman resonance condition. In the following part, we generalize the treatment to other velocities.

\section{Formulation of the effect as a phase shift}\label{sec:phase_shift}

\subsection{Case $\dLS=0$}

In order to obtain
a formula for any velocity, we calculate the phase shift that applies on
the interferometer arm that is not diffracted during the Raman transition. We study the following quantity for state
$\left|j\right\rangle$ ($j=1,2$): \begin{align} \delta\phi =
    \arg\left(\left\langle
    j\middle|U(\tau)\middle|j\right\rangle/\left\langle
    j\middle|U_0(\tau)\middle|j\right\rangle\right)\text{,}
\label{eq:dphi_diff_base} \end{align}which corresponds to the phase
difference between the cases with and without coupling.

Combining equations (\ref{eq:full_evol_no_ls}) and (\ref{eq:dphi_diff_base}),
we obtain the following formula for the additional phase shift induced by
the Raman coupling:
\begin{align}
    \delta\phi = 
    \arctan\left(\frac{\dD'}{\sqrt{1+\dD'^2}}
    \tan\left(\frac{\theta\sqrt{1+\dD'^2}}{2}\right)
    \right)-\frac{\dD'\theta}{2}\text{,}
    \label{eq:dphi_diff_no_ls}
\end{align}where $\theta=\Omega\tau$ is the pulse area and $\dD'=\dD/\Omega$ is the reduced Doppler detuning. 
If $\dD\rightarrow0$, we find as
expected that $\delta\phi\rightarrow0$, \textit{i.e} the coupling does not
induce an additional phase shift. On the other hand, for $\dD\gg\Omega$,
we develop equation (\ref{eq:dphi_diff_no_ls}) and obtain
$\delta\phi\approx\Omega^2\tau/(4\dD)$ which corresponds to the prediction
of the first order perturbation theory.

\begin{figure}
    \centering
    \includegraphics{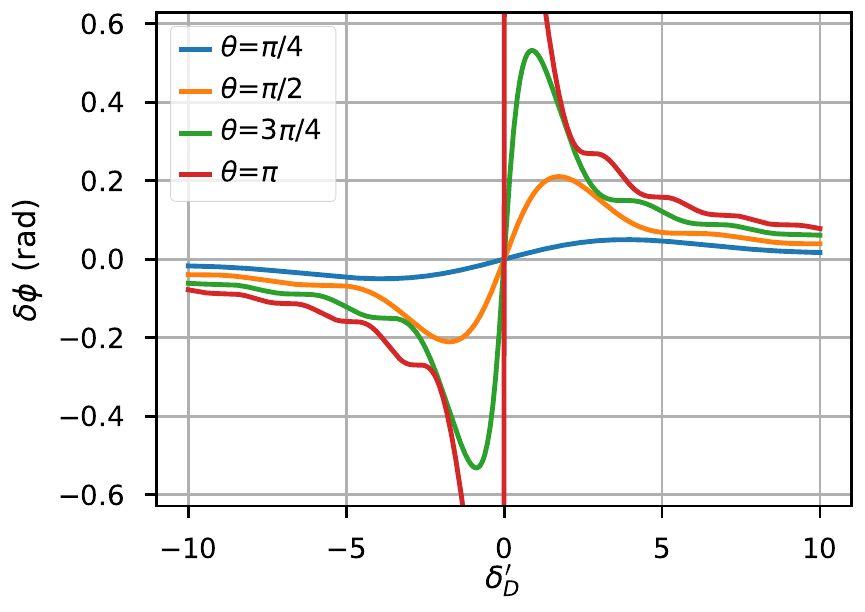}
    \caption{Plot of equation (\ref{eq:dphi_diff_no_ls}). Single-pulse phase shift
    with respect to $\dD'$ for different pulse areas.
    }
    \label{fig:dispersion_tls}
\end{figure}

We present the behavior of the intermediate case in figure
(\ref{fig:dispersion_tls}) where the phase shift is
plotted with respect to $\dD'$. We notice that the width of the
dispersion figure scales inversely with the pulse area. This can be
interpreted as an illustration of the Heisenberg principle: as the pulse
duration increases, its resonance frequency gets defined with a better
precision.

Now that we have treated the case of a single pulse, we can study the case of
an interferometric sequence. To this end we will continue considering the Ramsey-Bordé
sequence presented in figure (\ref{fig:ramsey_borde_dphi}). The perturbation
phase shift $\delta\phi$ is now given only by the $\arctan$ term of equation
(\ref{eq:dphi_diff_no_ls}) because the Doppler terms compensate each other.

The phase shift only applies on the wavepacket that remained in the same internal state during the Raman
transition. We
outline that the phase shift is the same in magnitude for the two states but changes 
sign.

\begin{figure}
    \centering
    \includegraphics{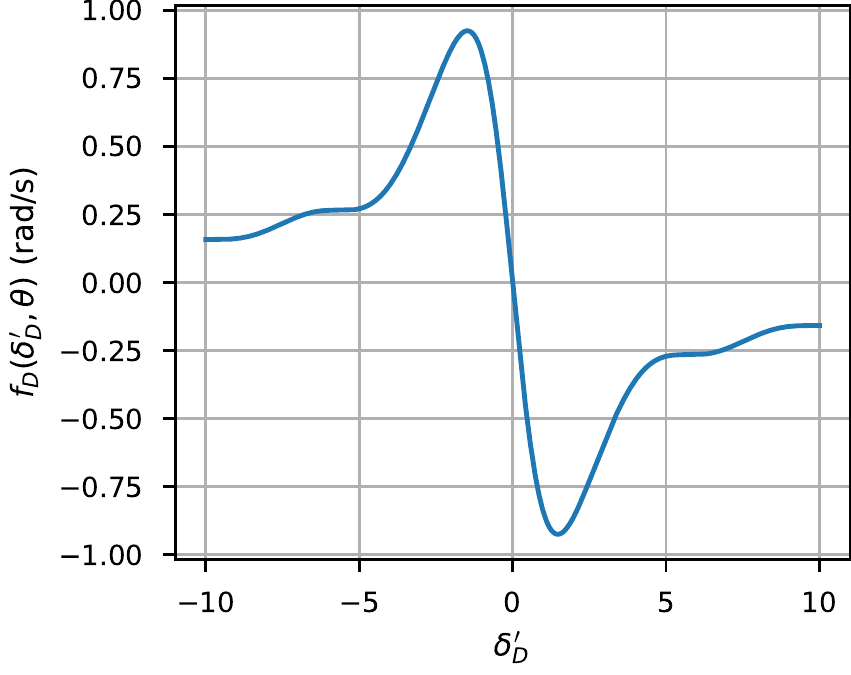}
    \caption{Plot of the function $f_D(\dD', \theta)$ (equation (\ref{eq:function_f})) for a Ramsey-Bordé interferometer with pulse area $\theta=\pi/2$.}
    \label{fig:theo_disp_interf}
\end{figure}

Moreover, we write the total phase shift at the output of the interferometer
$\Phi^{(B)}-\Phi^{(A)}$ where $(A)$ and $(B)$ are the arms displayed on
figure (\ref{fig:ramsey_borde_dphi}). We obtain that the contribution of this
parasitic effect writes:
\begin{align}
    \Delta\Phi = \delta\phi_4+\delta\phi_3-\delta\phi_2-\delta\phi_1
    \label{eq:contri_interf}
\end{align}
The sign of the contribution is the same for the two pulses of a single
pair, hence leads to an important non uniform phase shift that depends on the longitudinal velocity of the atoms. Besides, the effect is exalted by increasing the total
duration of the interferometer which is consistent with the $T$ scaling
obtained in equation (\ref{eq:displacement_rb}). More precisely, using the approximation of equation (\ref{eq:linear_rabi_var}), the formula (\ref{eq:contri_interf}) can be linearized with respect to the parameter $\beta T$. We obtain:
\begin{align}
    \Delta\Phi = \beta T f_D(\dD', \theta)\mathrm{,}
    \label{eq:linear_rb_dD}
\end{align}where the function $f_D$ is plotted in figure
(\ref{fig:theo_disp_interf}) and evaluates to:
\begin{align}
    f_D(\dD', \theta) = \frac{2\dD'\left(
    \sin\left(\theta\sqrt{1+\dD'^2}\right)
    -\theta\sqrt{1+\dD'^2}
    \right)
    }
    {
    \sqrt{1+\dD'^2}\left(2\dD'^2+\cos\left(\theta\sqrt{1+\dD'^2}\right)+1\right)
    }
    \text{.}
    \label{eq:function_f}
\end{align}This function behaves like a dispersion relation and displays its maximal sensitivity to $\dD'$ around zero, which corresponds to the usual operating conditions for interferometers. In order to reduce the sensitivity of the output phase of the interferometers to this effect, we act on the parameter $\beta$ and observe a reduction of the amplitude of the dispersion curve.

\subsection{Case $\dLS\neq0$}

\begin{figure}
    \centering
    \includegraphics{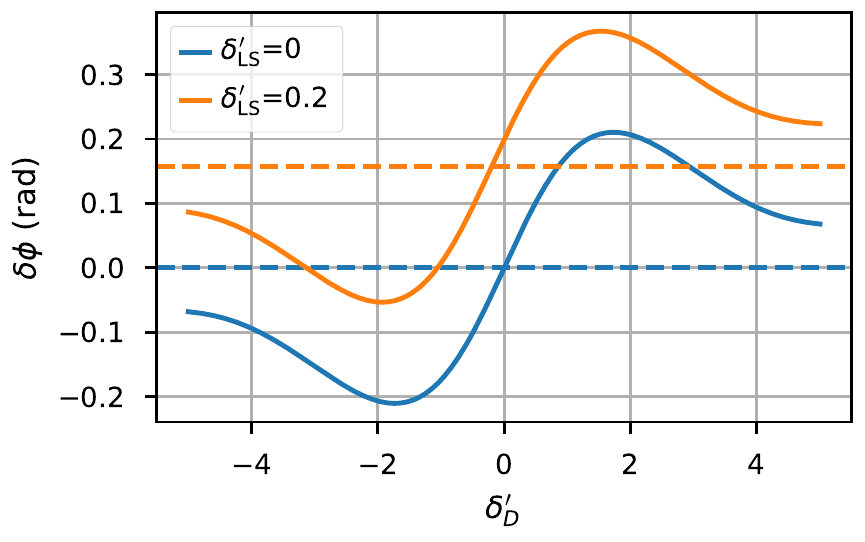}
    \caption{Single-pulse phase shift (equation (\ref{eq:dphi_diff_with_ls})) with respect to 
    $\dD'$ for a $\pi/2$ pulse and different light shift intensities.}
    \label{fig:dispersion_tls2}
    
\end{figure}

We now move to the case where the coupling induces an additional detuning
because of a differential light shift $\dLS\neq0$.
Equation (\ref{eq:dphi_diff_no_ls}) is modified and becomes:
\begin{align}
    \delta\phi = 
    \arctan\left(\frac{\delta'}{\sqrt{1+\delta'^2}}
    \tan\left(\frac{\theta\sqrt{1+\delta'^2}}{2}\right)
    \right)-\frac{\dD'\theta}{2}\text{,}
    \label{eq:dphi_diff_with_ls}
\end{align}where $\dLS'=\dLS/\Omega$ and $\delta'=\dD'+\dLS'$ is the total reduced detuning.
We represent in figure (\ref{fig:dispersion_tls2}) this equation with respect
to $\dD'$, for different $\dLS'$ and for a pulse area of $\pi/2$. We
observe both a vertical and horizontal displacement of the dispersion curve.
Indeed, the differential light shifts both modify the accumulated phase
between the two states and moves the resonance condition.

The question of the effects of level shifts induced by the coupling lasers has been addressed in the literature by analyzing a single pulse \cite{peters_high-precision_2001}, or globally using the sensitivity function of an interferometer \cite{gauguet_off-resonant_2008}. Using formula (\ref{eq:dphi_diff_with_ls}) in the limit of small differential light shift ($\dLS'\ll1$), we find that the phase shift for a single $\pi/2$ pulse amounts to $\delta\phi=\dLS'$ which is used in the above references. Our formula is valid beyond the limit of small shifts and also takes into account the effect of the Doppler shift. 

\begin{figure}
    \centering
    \includegraphics{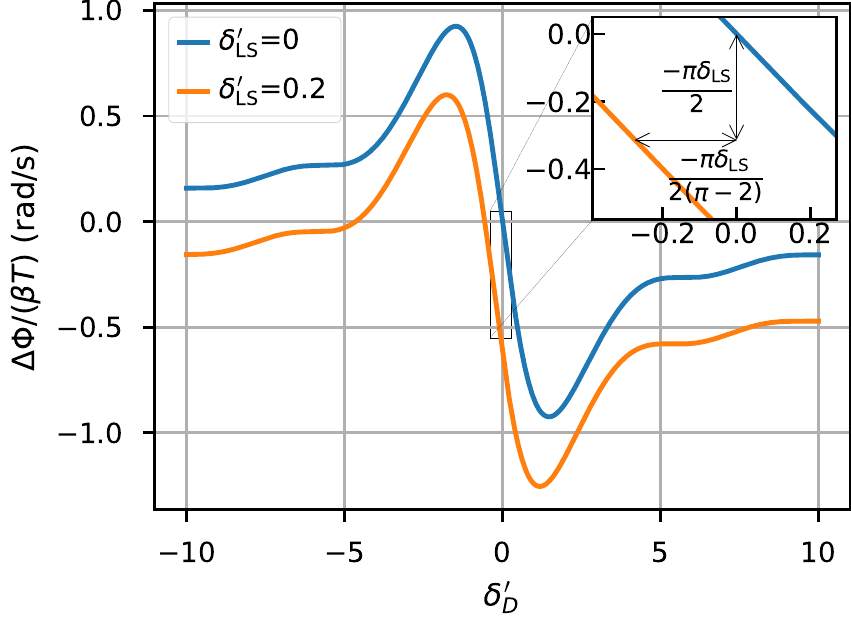}
    \caption{Effect on a Ramsey-Bordé interferometer in the presence of differential light shift $\dLS$ during the light pulse. Inset: horizontal and vertical displacement of the curve for $\dLS'\ll1$.}
    \label{fig:theo_disp_interf_ls}
\end{figure}

If the differential light shift $\dLS$ is constant during the interferometer, the phase shifts that it induces are cancelled. However, $\dLS$ is proportional to the lasers intensity like the Rabi frequency $\Omega$. Thus $\dLS$ varies and for the computations, we assume that it varies linearly with the same rate $\beta$. 
As a consequence, the dispersion curve of figure (\ref{fig:theo_disp_interf}) is modified. In particular, for $\dLS\ll\Omega$ it is displaced both horizontally and vertically similarly to the single-pulse case (figure (\ref{fig:dispersion_tls2})). We display these displacements of the curve in figure (\ref{fig:theo_disp_interf_ls}) and obtain that they write $-\pi\dLS'/2$ vertically and $-\pi\dLS'/(2(\pi-2))$ horizontally, if the interferometer pulses are $\pi/2$.

However, even with the presence of light shift, the amplitude of the dispersion curve remains proportional to the parameter $\beta$. We have developed an experimental method to reduce this parameter, based on the measurement of the dispersion curves. We present this protocol in section \ref{sec:observation}.

\section{Experimental setup}

\begin{figure}
    \centering
    \includegraphics[width=2.1in]{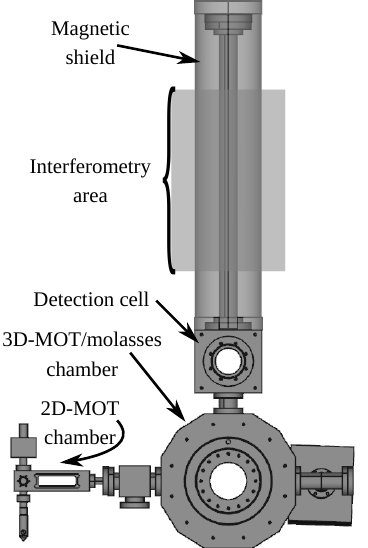}
    \caption{Schematic of the vacuum cell.}
    \label{fig:vacuum_cell}
\end{figure}

\begin{figure*}
    \centering
    \includegraphics{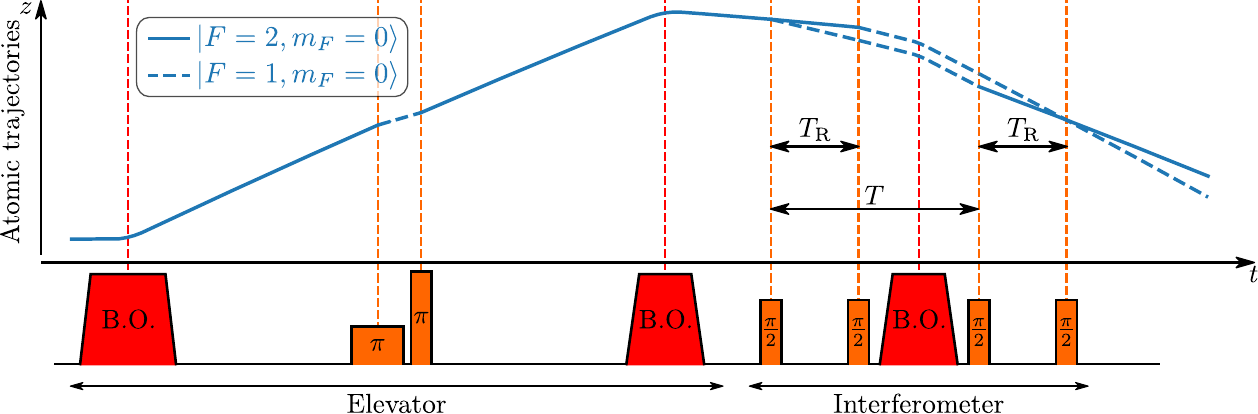}
    \caption{Experimental sequence to observe the dependence of the
    interferometer output with respect to the atoms initial velocity $\mdD$.
    Left: elevator sequence based on Bloch Oscillations, with the velocity and internal
    state selection based on two counterpropagating Raman $\pi$ pulses.
    Right: Ramsey-Bordé interferometer sequence.
    Blue line: Atom trajectory during an experimental sequence. Yellow lines: Raman pulses. Red areas: Bloch Oscillations pulses. The third Bloch Oscillations pulse, in the interferometer, is used to measure $h/m$. }
    \label{fig:the_trajectories}
\end{figure*}

The design of our vacuum chamber is shown in figure (\ref{fig:vacuum_cell}). In the lower area, we produce cold samples of $^{87}$Rb
atoms using conventional techniques (optical molasses and three-dimensional magneto-optical-trap). The atoms are cooled down to $\approx 4\ \mu$K.
Upon release, the size of the atomic cloud ($1/e^2$ radius) is 500 $\mu$m. The ground state of $^{87}$Rb has two hyperfine levels $\left|F=1\right\rangle$ and $\left|F=2\right\rangle$.

The interferometry sequence takes place in a 70 cm long tube  where the magnetic field is controlled within a few
percent. To transport atoms to the interferometry area, we apply
a Bloch Oscillations
(BOs) pulse~\cite{dahan_bloch_1996,
wilkinson_observation_1996} which transfers to the atoms 1300 atomic recoils
upwards. The cloud then reaches the interferometry area within 60 ms. We then
use a second 300 BOs pulse (600 atomic recoils) downwards 100 ms after the
first pulse to control the trajectory of the atomic cloud. We point out that
the timing and number of BOs parameters of the second pulse allows us to
control precisely this trajectory, and we refer to this technique as an atom
elevator\cite{clade_precise_2006}.

The preparation of the internal state is performed during the atom elevator as follows: at the output of the
optical molasses stage, the atoms are in $\left|F=2\right\rangle$ (a short
repump pulse removes all atoms from $\left|F=1\right\rangle$) distributed
among the five Zeeman sublevels of this hyperfine state. We apply a
counterpropagating (Doppler sensitive) Raman $\pi$-pulse set to be resonant
with the
$\left|F=2,m_F=0\right\rangle\rightarrow\left|F=1,m_F=0\right\rangle$
transition at the center of the cloud velocity distribution. The Raman lasers
are phase-locked, and their frequency difference and relative phase are
controlled using a radiofrequency generator based on a RedPitaya FPGA
device \cite{clade_improving_2017}. A blow-away pulse, resonant with the
$\left|F=2\right\rangle\rightarrow\left|F'=3\right\rangle$ optical transition,
removes atoms that did not perform the Raman transition. The atoms are then
placed back to $\left|F=2,m_F=0\right\rangle$ with a second
$\pi$ Raman pulse.
We control the width of the selected
velocity class through the duration of the first Raman pulse, maintaining the $\Omega\tau=\pi$ condition.
The
second Raman $\pi$ pulse is set at minimal duration, using the maximum available
laser power, in order to transfer back all the atoms selected by the first pulse.

In our setup, the waist of the Raman beams is $4.9\ $mm, the typical maximum intensity is $130\ $mW/cm$^2$ and the duration of a $\pi$ pulse is at least $60\ \mu$s which corresponds to a Rabi frequency of $16.6\ $kHz. The Raman lasers are blue-detuned with respect to
the one-photon transition by a large amount ($\approx 60\ $GHz). 
In this paper, we focus on the study of the effect on a single interferometer. For precision measurements, we compensate parasitic phase shifts using the k-reversal technique \cite{peters_high-precision_2001, clade_precise_2006}.

After this preparation stage, we apply the interferometric sequence. A schematic of
the total sequence is presented in figure (\ref{fig:the_trajectories}) which
represents the interferometer used to exhibit the effect (see section
\ref{sec:observation}).

The atomic state at the end of the interferometer is read by the detection
setup placed in-between the optical molasses cell and the interferometry
area. The free-falling atoms reach a system of resonant light sheets that
detects separately atoms in $\left|F=2\right\rangle$ and atoms in
$\left|F=1\right\rangle$ through a time-of-flight technique. We then
estimate the proportion of atoms in each state by collecting the
fluorescence the atoms emit as they cross the sheets onto large-area
photodiodes.

\section{Results}
\label{sec:observation}

\begin{figure}
    \centering
    \includegraphics{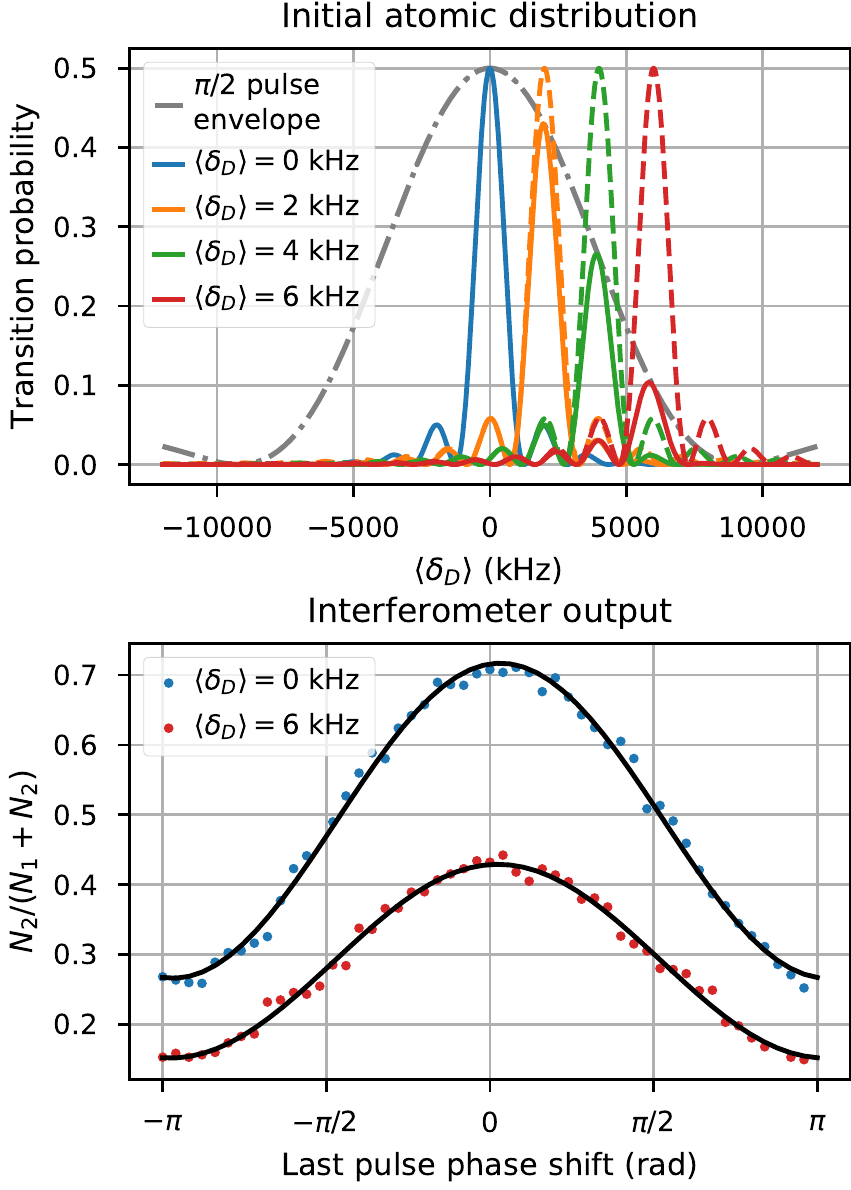}
    \caption{Top: velocity distribution at the beginning of the
    interferometer. Gray dotted line: first $\pi/2$ envelope. Dashed line:
    velocity distribution before the first pulse for different setpoints of
    the velocity selection stage. Solid lines: velocity distribution after
    the first pulse.\\ Bottom: proportion of atoms in
    $\left|F=2\right\rangle$ at the output of the interferometer with respect
    to the phase of the last $\pi/2$ pulse for two mean Doppler detunings.}
    \label{fig:raphael_spectra}
\end{figure}

In order to exhibit the effect, we implemented a Ramsey-Bordé
interferometer with $\tramsey=10\ $ms and $T=70\ $ms. The complete sequence is represented in figure
(\ref{fig:the_trajectories}) with the exception of the third Bloch Oscillations pulse, not implemented at this stage.

Treating
the Ramsey-Bordé interferometer at first approximation, the phase difference
at the output of the interferometer between its two arms is proportional to
the velocity acquired by the atoms between the first and third Raman
interferometer pulses. The phase shift presented in section \ref{sec:phase_shift} acts
as a perturbation of the differential velocity sensor signal.

In equation (\ref{eq:dphi_diff_with_ls}), both $\Omega$ and $\dLS$ depend on
the cloud expansion and are difficult parameters to control over the
interferometer. However, the parameter $\dD$ can be controlled by changing
the frequency difference between the Raman lasers at the velocity-selection
stage. Changing this parameter amounts to varying the mean Doppler detuning
$\mdD$ of the atomic cloud.

The top part of figure (\ref{fig:raphael_spectra}) outlines the details of
the technique. We scan $\mdD$ (dashed lines) without changing the Raman
lasers frequency for the interferometer (black dotted line). The effective
velocity distribution that contributes to the output of the
interferometer (solid lines) is then slightly different from the initial
distribution because of the interferometer pulse envelope.

We scan the phase
of the last Raman pulse with 50 points equally distributed on a $2\pi$
interval, and estimate the position of the center of the fringe (phase shift) with a sinusoidal fit.
This result is plotted on the bottom part of figure
(\ref{fig:raphael_spectra}) for $\mdD=0$ and $6\ $kHz. Because of the large
detuning at $\mdD=6\ $kHz, the contrast of the fringe is reduced by $\approx
40\%$.

\begin{figure}
    \centering
    \includegraphics{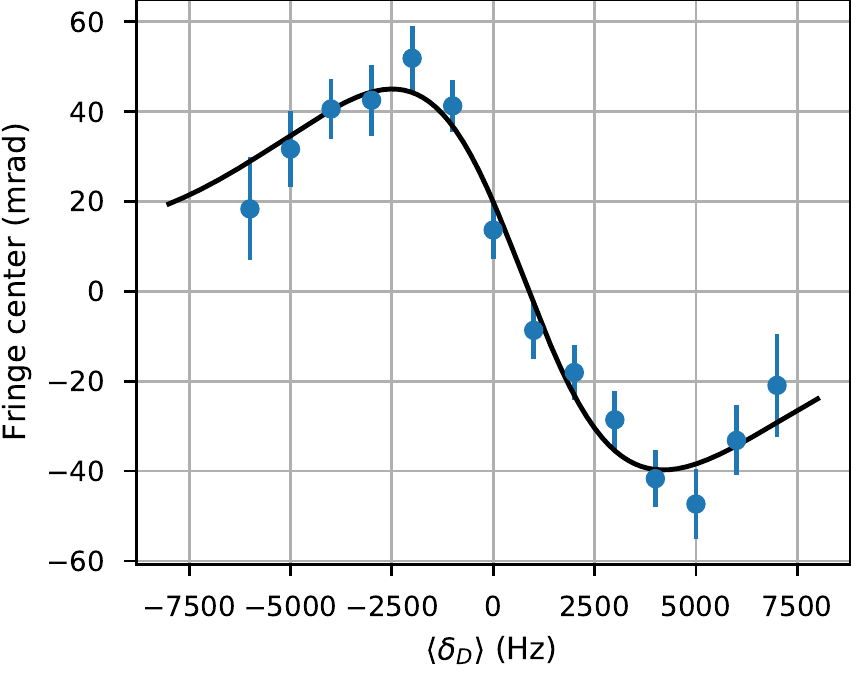}
    \caption{Plot of the fringe center with respect to the mean Doppler
    detuning of the atomic cloud. The origin of the vertical axis is set to
    the average value of the dataset for visibility. Blue points: experimental
    data. Black line: fit obtained using equation (\ref{eq:contri_interf}).}
    \label{fig:dispersion_interferometre}
\end{figure}

Using such a method, we obtain the phase at the output of the interferometer with
a typical precision of 10 mrad. In figure (\ref{fig:dispersion_interferometre}), we plot
the result of this measurement performed in the interval $[-6, 7]\ $kHz with a
step of $1\ $kHz. The measurement was repeated three times for each value
of $\mdD$.
The Rabi frequency of the interferometer pulses was 4.8\ kHz.

The results are consistent with the prediction of figure
(\ref{fig:theo_disp_interf}).
In particular, we observe that the zero of the curve is
slightly shifted because of the light shift.

The amplitude of the phase shift ($\approx 80\ $mrad) is important compared
to the aimed accuracy of this atom interferometer. For instance, the required precision for the
measurement of $h/m$ at the level of 0.1 ppb stands below the mrad level. Moreover, this effect depends on the longitudinal velocity of the
atoms so that a careful estimation requires precise knowledge of the
velocity distribution. However, even if one implements a control of the atomic
velocity distribution, correlations between the atomic velocities and the efficiency of the 
detection system make a more accurate evaluation of this effect challenging \cite{farah_effective_2014}. Hence, we developed
an experimental method to reduce the effect.

\begin{figure}
    \centering
    \includegraphics{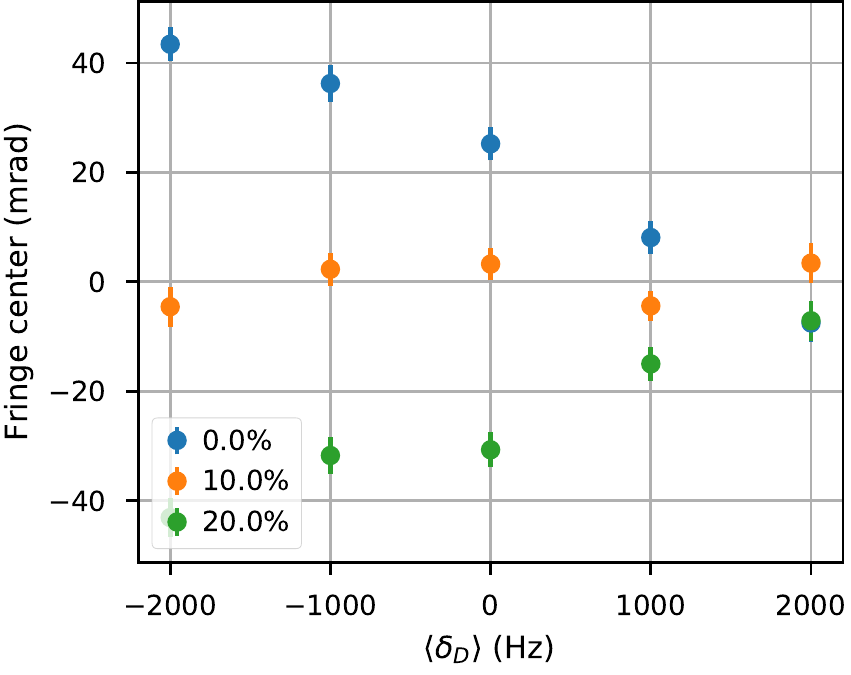}
    \caption{Fringe center of the interferometer with respect to the mean
    Doppler detuning of the atomic cloud for variations of the laser power
    between the first and the last pulse. For this study, the Rabi frequency
    was 8 kHz. The origin of the vertical axis is set to
    the average value of the dataset for visibility.}
    \label{fig:compensation_expansion}
\end{figure}

Because the parasitic phase shift is induced by the variation of perceived
power, we increase the power of the Raman lasers, linearly with time, during
the interferometer. We chose to apply a linear ramp because in our experiment
the interferometry sequence begins about $170\ $ms after the release from the
optical molasses phase: the transverse position of the atoms depends only on
their velocity and we expect a quasi-linear variation of the average
intensity perceived by the cloud with time.

The idea of the technique is to apply different laser power ramps on the Raman
lasers, and scan the mean Doppler detuning of the atomic distribution at the
input of the interferometer. As the perceived intensity is made more
homogeneous for all the interferometer pulses, the dependency of the output
phase with the Doppler shift is reduced. In particular, if the Rabi frequency
$\Omega$ and the differential light shift $\dLS$ are the same for the four
pulses, the phase shift of equation (\ref{eq:dphi_diff_with_ls}) is constant
and is cancelled according to equation (\ref{eq:contri_interf}).

For the measurement of $h/m$, we insert in-between the Ramsey
sequence a pulse of 500 BOs and measure the velocity the atoms
acquired \cite{bouchendira_new_2011}.  BOs is an adiabatic process whose
survival probability (\textit{i.e} the probability to absorb two atomic
recoils) depends sharply on the intensity and can be modeled by a step
function \cite{bade_observation_2018}: atoms that perceive an intensity below
a given threshold do not perform the BO and atoms above perform it. Thus, the
BOs pulse performs a transverse selection of the atoms as those at the edge
of the beams are not accelerated. The effective transverse velocity
distribution that contributes to the interferometer signal depends on this BO
pulse.

Hence, we performed the study with the BO pulse, \textit{i.e} in the
condition of the measurement of $h/m$ for different Raman power ramps: no power ramp
and linear ramps that correspond to a power increase of 10\% and 20\% from
the first to the last pulse of the interferometer. The result of this study
is displayed in figure (\ref{fig:compensation_expansion}). We observe that
for the largest power ramp, the sign of the dependency of the phase with the
mean Doppler detuning is changed, which means that the average intensity
perceived by the atomic cloud increases along the interferometer instead of
decreasing.

Finally for a 10\% increase of the power at the last pulse, the effect is
almost cancelled. We do not observe a perfect cancellation because the method
of the power ramps is based on reducing the variation of the average intensity perceived by the atoms. Even with this power ramp,
the atoms with zero transverse velocity perceive an increasing intensity along
the interferometer and the atoms with the largest transverse velocity still
perceive a decreasing intensity but with a smaller amplitude than without the
power ramp. 

Because of this limitation, we estimate that we reduced the sensitivity of
the interferometer to the initial velocity distribution by 80\%, which
corresponds to the reduction of the amplitude of the curves displayed on
figure (\ref{fig:compensation_expansion}).

These measurements outline that the sensitivity of the interferometer to the velocity distribution originates from the laser intensity variation of the Raman pulses. They also show that it is possible to reduce the sensitivity by applying an adequate power ramp. We highlight the fact that our method does not require \textit{a priori} information on the velocity distribution of the atoms and that the power ramp is experimentally calibrated.

\section{Conclusion}

In this paper, we have shown that during a Raman transition, the wavepacket that remains in the initial state undergoes a displacement. Because of variations of
the Rabi frequency of the pulses, this displacement is not cancelled at the
output of the interferometer which induces a sensitivity to the initial
velocity distribution. By measuring the phase of the interferometer with respect to the initial velocity, we identify the dispersive signature of this effect. We have obtained an analytical formula of the phase of the interferometer that corroborates experimental observations. These general formulas can be adapted to any interferometer geometries based on Raman transitions.

The dispersive behavior as a function of the initial velocity is crucial to experimentally compensate the variation of the laser intensity perceived by the atoms during the interferometer. This protocol can be applied on particular interferometer configurations such as the one we use for the measurement of $h/m$.

Although we focused our treatment on Raman transitions, the model we have developed in this paper for velocity-dependent phase shift could be adapted to Bragg diffraction. We also believe that this effect should be accurately evaluated in experiments aiming at testing the equivalence principle with dual-species interferometer \cite{dimopoulos2007,Bonnin2013,schlippert2014,rosi2017,asenbaum2020}  and our approach would help for seeking for a signature enabling a better compensation of the systematic bias resulting from the variation of the laser intensity during the light-pulse sequence.

\bibliographystyle{apsrev4-1}
\bibliography{papier_light_shift}

\end{document}